# Wavelet analysis of non-Gaussian anisotropies from primordial voids in simulated maps of the Cosmic Microwave Background

Fabio Noviello[a, 1]

[a]*National University of Ireland Maynooth, Maynooth, Co. Kildare, Ireland*

*E-mail address*: fabio.noviello@ias.u-psud.fr

*Phone*: +33-(0)169858763

**Abstract**

Phase transitions taking place during the inflationary epoch give rise to bubbles of true vacuum embedded in the false vacuum. These bubbles can imprint a distinctive signal on the Cosmic Microwave Background (CMB). We evaluate the feasibility of detecting these signatures with wavelets in CMB maps, such as those that will be made available by the European Space Agency's (ESA) Planck mission.

## 1. Introduction

The standard inflationary model predicts that Gaussian quantum fluctuations of the inflationary field (inflaton) will be imprinted on the surrounding matter distribution, and hence on the Cosmic Microwave Background, through the decay of the inflaton into particles and radiation. Inflationary scenarios with first-order phase transitions suggest that the system (our universe) tends to regain its equilibrium by enucleating bubbles of true vacuum (TV) by quantum tunnelling through the potential barrier existing between the false vacuum (FV) and TV phases. The perturbations induced in the primordial

---

[1] Permanent address: *Institut d'Astrophysique Spatiale, CNRS & Univ. Paris-Sud, UMR 8617, 91405 Orsay Cedex, France*

plasma by these bubbles can leave an imprint on the CMB at decoupling (Amendola et al., 1998; Griffiths et al. 2003).

In this paper we explore the possibility that the imprints of these primordial bubbles could, in principle, be detected with wavelets on the maps of the CMB that will be produced by the ESA Planck mission in the near future (Tauber, 2004). Section 2 outlines the cosmological interest of the primordial bubbles and describes their signature on the CMB, while in Section 3 we give a brief overview of wavelets. In Section 4 we discuss the simulated map-making procedure used in this work and then proceed to Section 5 we present the results of our analyses. Finally, in Section 6, we draw our conclusions.

## 2. Physics of void signatures on the Cosmic Microwave Background.

The bubbly perturbations produced by TV bubbles generate an expanding spherical wave that gives rise to a number of weak (hot and cold) rings with radial dimensions reaching up to those of the sound horizon at decoupling ($\cong 0.6$ deg). At decoupling, a strong central negative (cold) spot is left surrounded by traces of the outward expanding rings. The cold spot is due to the fact that the void has a slightly overcomoving expansion rate (Baccigalupi et al., 1997; Amendola et al., 1999). Therefore, photons crossing the void will, essentially, experience a higher redshift than those in the surrounding medium. The physical size of a void grows as $t^{4/5}$ while the scale factor of surrounding space in the matter dominated era increases as $t^{2/3}$. Therefore, at decoupling, a void was smaller by a factor of four than today.

The amplitude of the temperature anisotropies due to the voids, can be estimated as (Amendola et al., 1998)

$$\left(\frac{\Delta T}{T}\right)_V \cong -\delta \left(\frac{R_V^{dec}}{R_H^{dec}}\right)^2 \quad (1)$$

where $R_V^{dec}$ is the void radius at decoupling while $R_H^{dec}$ is the Hubble radius, also at decoupling ($\approx 100h^{-1}$ Mpc). The density contrast $\delta$ is defined as

$$\delta = \frac{\rho_C}{\rho_\infty} \quad (2)$$

where $\rho_C$ is the matter density at the void centre, while $\rho_\infty$ represents the average matter density outside the void. Primordial voids have also been proposed as an alternative (or complementary) cosmological structure formation mechanism as, for instance, in (Occhionero et al., 1997). Seeds for the formation of the first generation of galactic objects might be formed at the void shells by shock fronts created by the compressed surrounding medium during their expansion. The thin shell contains matter that is swept up during void expansion. This *might* contribute to explain the "bubbly" large-scale structure observed in galaxy surveys. The discovery of the imprint of voids on the CMB would therefore be doubly interesting, mainly as a probe of the validity of the inflationary paradigm but also from the point of view of structure formation mechanisms.

We must now address the issue of the fraction of space filled by *observable* voids. Although different models exists, a general conclusion derived from (Coleman, 1977) is that the bubble sizes at decoupling follow a power-law distribution, such as

$$n_B(L) = \left(\frac{L_{\max}}{L}\right)^p \qquad (3)$$

where $n_B(L)$ is the bubble density per unit volume and $L$ is the comoving radius of the bubble. The quantities $L_{\max}$ and $p$ are parameters whose values are model-dependent. For instance (Occhionero et al., 1997), within the framework of two-field inflation with quadratic gravity, find values of $L_{\max} \cong 30 h^{-1}$ Mpc and $p \in [3,10]$. The void radii cutoffs can be chosen to agree with large scale structure surveys, such as in (Mathis et al., 2004).

Regarding the direct detection of voids from the CMB power spectrum, the most interesting case is the one in which the imprint of the primordial voids exists but their presence cannot be established by analysing the power spectrum. (Baccigalupi and Perrotta, 2000) find the following empirical relation

$$\delta_{ps} = \sqrt{10^{-6} \left(\frac{20 h^{-1}}{R_v^{dec}}\right)^7 \left(\frac{0.5}{F}\right)}. \qquad (4)$$

where $F$ is the fraction of space filled by voids at decoupling. Voids with underdensities $\delta < \delta_{ps}$ will not be discernible from a power spectrum analysis while, in the opposite case, observable peaks should develop in the CMB power spectrum at the void scales.

The signal imprinted by voids on the LSS is clearly of a non-Gaussian nature, since it will identify preferential scales where the CMB anisotropies will exhibit a correlation due to the voids. This implies that scale-dependent data analysis techniques would be particularly suitable for their detection.

## 3. Wavelets

In this section, meant as a brief description of wavelets, designed to aid the comprehension of our results presented in the next section, we follow mainly (Dremlin et al., 2001). Considering a discrete set of coordinate labels (corresponding to an experimental data set), we can define the (discrete) *scaling function* in terms of contracted and shifted versions of itself as

$$\varphi(x) = \sqrt{2} \sum_{k=0}^{2M-1} h_k \, \varphi(2x-k) \qquad (5)$$

where $\varphi(2x-k)$ is a contracted version of $\varphi(x)$ shifted along the *x*-axis by an integer step $k$ and factored by the scaling coefficient $h_k$. The integer $M$ defines the number of translations and contractions, which are equal to $2M$. Different values of $M$ define different wavelets within a specific family. The scaling coefficients $h_k$ are defined as

$$h_k = \sqrt{2} \int dx \, \varphi(x) \varphi^*(2x-k) \qquad (6)$$

where the asterisk denotes complex conjugation. The *mother wavelet* is then built from the scaling function through the following equation

$$\psi(x) = \sqrt{2} \sum_{k=0}^{2M-1} g_k \, \varphi(2x-k) \qquad (7)$$

where the scaling coefficients are given by

$$g_k = (-1)^k \, h_{2M-1-k}. \qquad (8)$$

It is essential to point out that, in general, the scaling and wavelet functions have no explicit analytic form but, rather, are defined through Equations 5 and 7. The contracted and translated versions of these functions are

$$\varphi_{j,k}(x) = 2^{-\frac{j}{2}} \varphi(2^{-j} x - k) \qquad (9)$$

$$\psi_{j,k}(x) = 2^{-\frac{j}{2}} \psi(2^{-j} x - k). \qquad (10)$$

Here, the index $j$ denotes the contraction scale. From the point of view of data analysis, it is the level of detail at which the data is being analysed. Here we are following the convention according to which increasing values of the index $j$ denote a progressively finer sampling of the data, as in (Fang and Thews, 1998). The set of functions $\{\varphi_{j,k}, \psi_{j,k}\}$, with $j \in [0, \infty)$ and $k \in (-\infty, \infty)$ forms a complete orthogonal basis (Daubechies, 1992).

$N$-dimensional scaling functions and wavelets can be built by taking the tensor product of one-dimensional bases. In the two-dimensional case we have

$$\Phi_{j_1, k_1; j_2, k_2}(x_1, x_2) = \varphi_{j_1, k_1}(x_1) \varphi_{j_2, k_2}(x_2) \qquad (11)$$

$$\Psi_{j_1, k_1; j_2, k_2}(x_1, x_2) = \psi_{j_1, k_1}(x_1) \psi_{j_2, k_2}(x_2) \qquad (12)$$

where $x_1$ and $x_2$ designate variables along the two dimensions, which are scaled independently from one another. By using orthogonality relations between scaling functions and wavelets and imposing conditions on the number of vanishing moments, one can construct different wavelets and wavelet families, such as the Haar wavelet used in our analysis and shown in Figure 1.

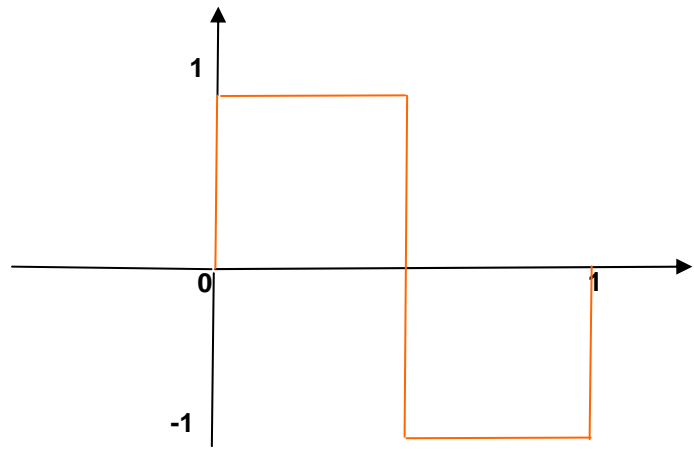

**Figure 1**
The Haar wavelet.

Any function $f(x) \in L^2(R)$ can be expanded in a *Discrete Wavelet Transform* (DWT). At the *n*-th resolution level, the function will be represented by the following series

$$f(x) = \sum_{k=-\infty}^{+\infty} s_{j_n,k} \varphi_{j_n,k}(x) + \sum_{j=j_n}^{\infty} \sum_{k=-\infty}^{+\infty} d_{j,k} \psi_{j,k}(x) \qquad (13)$$

where $j_n$ denotes a specific scale. In principle, the scaling and wavelet function expansion coefficients at scale $j$ can be computed as

$$s_{j,k} = \int dx\, f(x) \varphi_{j,k}(x) \qquad (14)$$

$$d_{j,k} = \int dx\, f(x) \psi_{j,k}(x). \qquad (15)$$

From a data analysis viewpoint, if a one-dimensional data sample comprises $N = 2^J$ elements then the indices will vary in the following ranges: $j \in [0, J-1]$ and $k \in [0, 2^J - 1]$. A more practical way to proceed, used in our work, is to apply *pyramidal algorithms* for computing the scaling and wavelet expansion coefficients (Mallat, 1998). The only input one needs to start the algorithm is the value of the first scaling coefficient, $s_{0,k}$. The $s_{j_n,k}$ (*scaling*) coefficients describe the *average* value of the data, at a specific scale $j_n + 1$, for a wavelet-dependent number of adjacent data bins. The $d_{j_n,k}$ (*detail*) coefficients, measure the *differences* between a number of adjacent bins at scale $j_n + 1$. In Figure 2 we show an idealised example of multiresolution analysis applied to a histogram with four data bins.

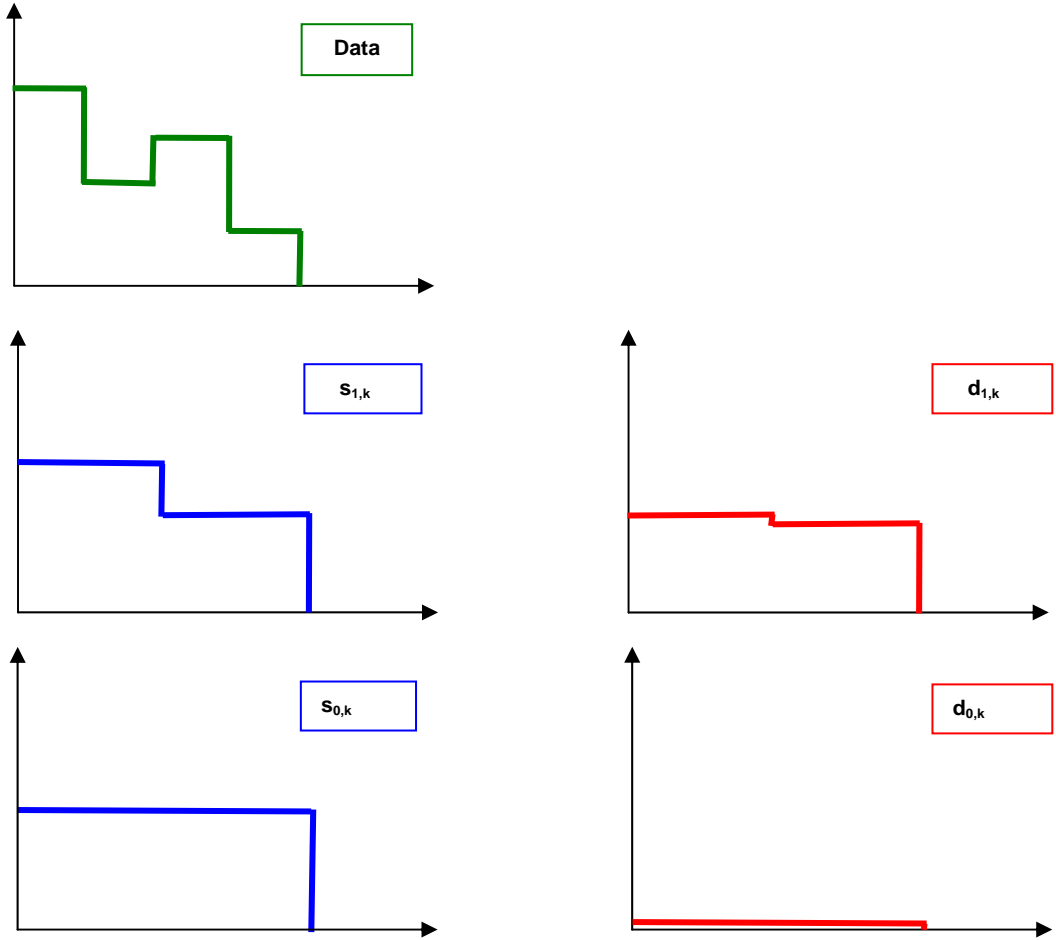

**Figure 2**
Idealised discrete wavelet decomposition of the data displayed in the histogram (*green*). The $s_{j,k}$ coefficients (*blue*) probe the mean values at the scale $j + 1$, while the wavelet expansion coefficients (*red*) investigate the details of the data fluctuations, also at the scale $j + 1$. The specific decomposition in this figure corresponds to the Haar wavelet (apart from normalisations).

## 4. Simulation of CMB maps

### 4.1. Gaussian realisations

The primary templates for our work are maps of the CMB sky with Gaussian anisotropies generated with CMBMAP (Muciaccia et al., 1997). This code generates full-sky maps of CMB anisotropies, using an *Equidistant Cylindrical Projection* (ECP). In an ECP, distances along a parallel are conserved and the polar regions are highly distorted. The input power spectrum for CMBMAP was generated using CMBFAST (Seljak & Zaldarriaga, 1996). We used the WMAP (first year) best-fit cosmological parameters (Bennet et al., 2003) since the three-year data was not yet available when this work was begun. The maps being distorted in the polar regions, we extracted 50 smaller square Gaussian maps from the equatorial region, each map having a size of 22.5 deg$^2$ and a resolution of 5.27 arcminutes. These smaller maps can be considered as being approximately planar and, consequently, suitable for a DWT analysis. Each one was derived from a different Gaussian realisation of the full-sky map, but all were extracted from the same region of the sky. An example of a small Gaussian map is shown in Figure 3.

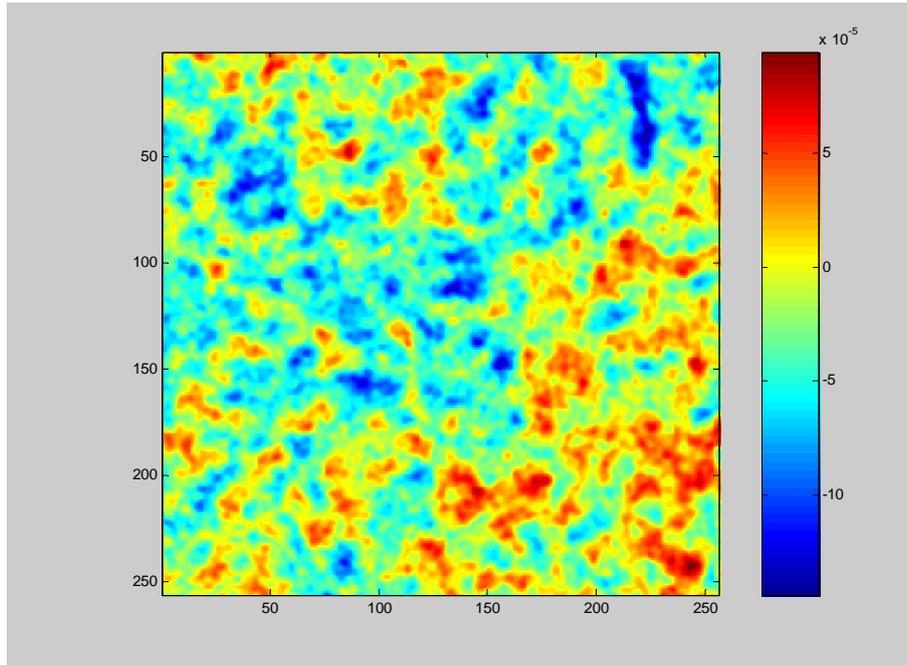

**Figure 3**
Gaussian CMB anisotropy map. Field of view = 22.5 deg$^2$, resolution = 5.27 arcminutes. The scales on the coordinate axes denote the pixel numbers, while the colourbar is relative to the temperature anisotropy $\Delta T / T_0$ (in units of $10^{-5}$).

### 4.2. Maps with imprints of inflationary voids

As for the bubbles themselves, only the central voids were considered in our simulations, and not the surrounding acoustic rings. This because the void signal is much stronger than the rings' and can therefore be considered as the primary tracer of their eventual presence, as well as to simplify computations. This approach is consistent, for instance, with (Griffiths et al. 2003). We considered voids with radii of $R_V^{dec} = 7.5 h^{-1}$ Mpc at decoupling, falling at the lower end of the range of values predicted by theory (Amendola et al., 1999). Smaller voids are erased by matter inflow. At the distance of the decoupling surface (14 Gpc) the voids have an angular size of about 5.3 arcmin. Therefore, in our simulations, each void corresponds to a map pixel.

Due to their overcomoving expansion rate, the same voids will have radii of $R_V^0 = 15h^{-1}$ Mpc *today*, this value being consistent with the sizes of voids observed in large-scale structure surveys.

In our simulations we considered four different filling fractions *at the present epoch*, their values being $F$ = 40%, 20%, 2% and 1%. The highest fraction corresponds to the case in which all of the currently observed voids in the large-scale distribution of matter can be traced directly to the primordial ones. In this scenario they would be the primary seeds of large-scale structure formation. The other filling fractions correspond to models with progressively decreasing values of the parameter *p* in Equation 3 (with fixed $L_{max}$). As the value of *p* decreases there is an increasing number of smaller voids, which are cancelled by material inflow and a smaller number of surviving voids that can be detected in the CMB. The number of voids was estimated by filling a cube with sides of $3000h^{-1}$ Mpc with non-intersecting spheres with radii equal to $R_v^0$. The pixels representing the voids were distributed randomly on grids having the same size and resolution of the previously simulated ΛCDM[2] maps. Multiple enucleation of voids from a single point (pixel) is avoided, so the number of voids on a grid goes from $N_V \cong 4800$ in the $F$ = 40% case to $N_V \cong 134$ for $F$ =1%. For each filling fraction two different values of the underdensity, $\delta = 1\%$ and $\delta = 0.5\%$, were used. The resultant values of the induced temperature anisotropies are $\Delta T/T_0$ = -5.625×10$^{-5}$ ($\delta = 1\%$) and $\Delta T/T_0$ = -2.825×10$^{-5}$ ($\delta = 0.5\%$). We simulated 50 independent realisations for each different combination of $F$ and $\delta$, for a total of 400 void-only "maps". These were then superimposed on the previous set of ΛCDM realisations. This direct superposition is legitimate since both the Gaussian and bubble perturbations are linear. Two examples of voids + ΛCDM maps, from opposite ends of parameter space, are shown in Figures 4 and 5. It is worth noticing that a void can show up as either a cold or a hot spot on the CMB. Also, although present, voids are not visible in the map shown in Figure 5

---

[2] From now on the label 'ΛCDM' will characterise our Gaussian maps, as opposed to non-Gaussian ones.

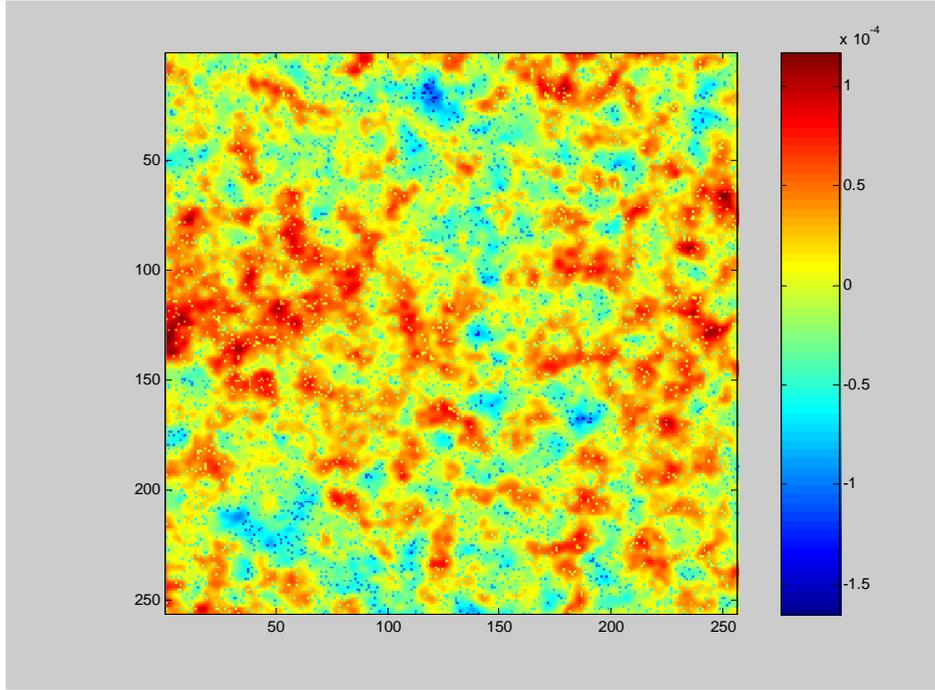

**Figure 4**

ΛCDM + voids map with filling fraction $F = 40\%$ and void underdensity $\delta = 1\%$. Field of view = 22.5 deg$^2$, resolution = 5.27 arcminutes. The scales on the coordinate axes denote the pixel numbers, while the colourbar is relative to the temperature anisotropy $\Delta T/T_0$ (in units of $10^{-4}$).

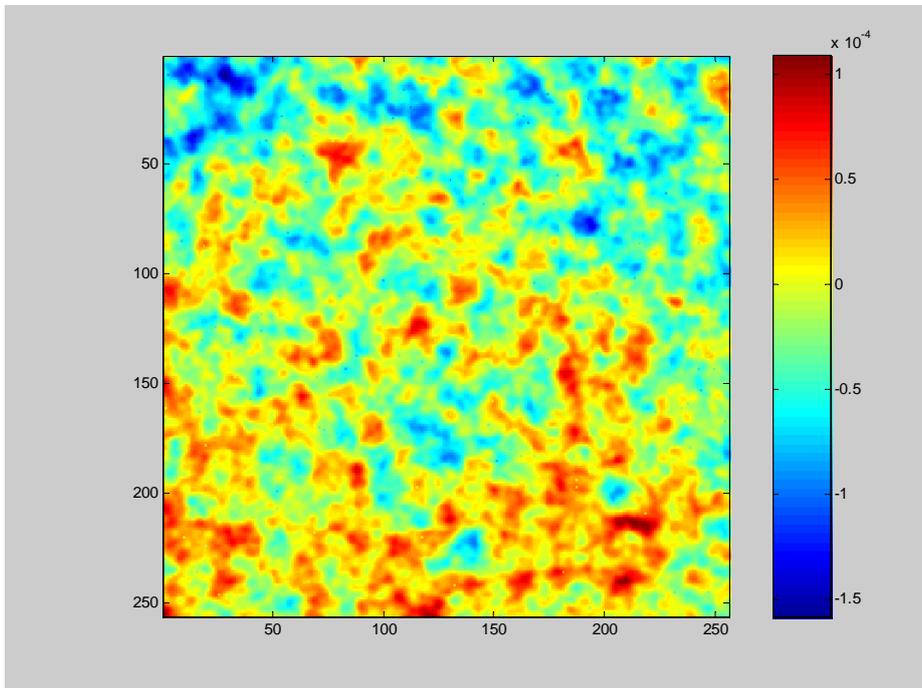

**Figure 5**

ΛCDM + voids map with filling fraction $F = 1\%$ and void underdensity $\delta = 0.5\%$. FOV = 22.5 deg$^2$, resolution = 5.27 arcminutes. The scales on the coordinate axes denote the pixel numbers, while the colourbar is relative to the temperature anisotropy $\Delta T/T_0$ (in units of $10^{-4}$).

## 5. Analysis of CMB maps

Wavelet analysis of the CMB has been applied to both simulated and experimental data. For instance (this list is not exhaustive), from a theoretical standpoint it has been used to detect the Kaiser-Stebbins effect, produced by cosmic strings (Hobson et al., 1999) and for the denoising of simulated maps (Sanz et al., 1999). Spherical Haar wavelet analysis of WMAP one-year data detected a cold spot at southern galactic latitudes with an angular size of about 10 degrees. After excluding systematic causes, foreground sources and the Sunyaev-Zel'dovich effect, (Cruz et al, 2005) conclude that this spot must be intrinsic to the CMB. This result was confirmed by a Spherical Mexican Hat wavelet analysis of the WMAP three-year data (Cruz et al, 2007). The origin of the spot is currently open to debate. The three-point collapsed function and normalised bispectrum are applied to the detection of primordial voids by (Corasaniti et al., 2001). The work presented in this paper is the first in which wavelets are applied to their detection. The wavelet software used in the following analysis is based on *Numerical Recipes* Fortran routines (Press, 1992) while the remaining code is custom-made.

### 5.1. Choice of wavelet bases and statistical estimators

According to (Hobson et al., 1999) and (Dremlin et al., 2001) the *Maximum Entropy method* can be used to determine the optimal wavelet basis for the recovery of a specific signal. One begins by computing the two-dimensional wavelet expansion coefficients $d_{j,k_1 k_2}$, with $j = j_1 = j_2$, on a single sample map[3]. Next, for each value of $j$, the "normalised" coefficients are given by

$$p_{j,k_1,k_2} = \frac{d^2_{k_1,k_2}}{\sum_{k_1,k_2} d^2_{k_1,k_2}}. \qquad (16)$$

Here the sum in the denominator extends over all values of $k_1$, $k_2$, relative to the scale $j$. The entropy of the normalised coefficients is then defined as

$$S = -\frac{1}{\log N} \sum_{k_1,k_2} p_{j,k_1,k_2} \log\left(p_{j,k_1 k_2}\right) \qquad (17)$$

---

[3] Generally, $j_1 \neq j_2$. Here we are restricting the description of the method to our specific case.

where *N* is the total number of coefficients at the scale *j*. Since the DWT is a linear operator, one might expect a CMB map with Gaussian anisotropies to translate into a Gaussian distribution of the wavelet expansion coefficients. Moreover, for a Gaussian CMB map with structure on a wide range of scales we could expect that, at each scale, all the coefficients would be required, with *roughly* equal amplitudes, to represent the map on that scale. Therefore the entropy of the normalised wavelet coefficients should be close to its maximum value of unity. On the other hand, if there exists a wavelet basis that resembles the non-Gaussian features of a map at some scale, many of the coefficients will have low values (close to zero) with just a few larger coefficients representing the non-Gaussian features. Consequently, regardless of the name of the method, we are actually trying to *minimise* the entropy when trying to detect non-Gaussian signatures.

The sharp cutoff originating from the edges of our maps can give rise to artificial non-Gaussian features. Since the support of the wavelets may extend over several pixels, it is important to identify the latter. To do this, we created an array consisting of zeroes, except for the edges, which were made up of ones. "Contaminated" pixels were identified and discarded in our subsequent analysis.

We applied the entropy method on the wavelet expansion coefficients extracted from three test maps: a Gaussian one, together with two voids + ΛCDM maps with void parameters given by ($F = 40\%$, $\delta = 1\%$) and ($F = 40\%$, $\delta = 1\%$). The Haar wavelet, ten bases from the Daubechies family and five Coiflet wavelets were used

The lowest entropy at the pixel (void) scale was found to be the one related to the Daubechies 4 wavelet. Also, although its entropy at the $j = 7$ scale was not the lowest, we decided to experiment with the Haar wavelet. By taking into account its shape (see Figure 1) we considered that it might be useful in detecting "square" signals, such as voids represented as pixels. Since, in our subsequent analysis, the Daubechies 4 wavelet led to an unsatisfactory outcome, in this paper we focus on reporting the results obtained with the Haar wavelet. This also implies that the entropy method can be used as a guide for choosing an optimal wavelet for analysing a specific signal, but is not necessarily fail-safe.

The first statistical estimator we used is the *skewness* , defined as

$$Skewness(x_1, x_2, ...x_N) = \frac{1}{N} \sum_{i=1,N} \left[ \frac{x_i - \langle x \rangle}{\sigma} \right]^3 . \qquad (18)$$

Here $\langle x \rangle$ is the mean of the data set while $\sigma$ is its standard deviation. The skewness is a measure of the asymmetry of a distribution around its mean. Its value for a Gaussian distribution is zero, while a non-zero value of implies that the underlying distribution has an asymmetric tail. It can be expected that a finite date set will give non-zero values for Equation 18, even if the underlying population has a Gaussian distribution, due to the limited sample. We also employed the *excess kurtosis*, defined as

$$Kurtosis(x_1, x_2, ...x_n) = \left\{ \frac{1}{N} \sum_{i=1,N} \left[ \frac{x_i - \langle x \rangle}{\sigma} \right]^4 \right\} - 3 . \qquad (19)$$

The excess kurtosis is a measure of the relative peakedness of a distribution with respect to a Gaussian one. For a normal distribution, its numerical value is zero. The skewness and kurtosis are both dimensionless quantities.

The wavelet analysis software used in this analysis work is based on *Numerical Recipes* routines (Press, 1992) while the remaining code is custom-made.

### 5.2. Results for the ideal case with no instrumental noise

The skewness and kurtosis of the wavelet expansion coefficients were computed for the complete set of simulated maps discussed in Section 4, for scales not contaminated by edge effects. The results obtained are given in Table 1.

| $F$ (%) | $\delta$ (%) | Scale $j$ | Skewness | $\sigma_S$ | Kurtosis | $\sigma_K$ |
|---|---|---|---|---|---|---|
| 0 | 0 | 4 | $-2.4 \times 10^{-2}$ | $3.6 \times 10^{-1}$ | $-1.5 \times 10^{-1}$ | $4.8 \times 10^{-1}$ |
| " | " | 5 | $9.2 \times 10^{-3}$ | $7.5 \times 10^{-2}$ | $-5.6 \times 10^{-2}$ | $1.8 \times 10^{-1}$ |
| " | " | 6 | $1.6 \times 10^{-3}$ | $4.1 \times 10^{-2}$ | $8.5 \times 10^{-3}$ | $1.0 \times 10^{-1}$ |
| " | " | 7 | $5.0 \times 10^{-3}$ | $2.0 \times 10^{-2}$ | $6.2 \times 10^{-3}$ | $4.0 \times 10^{-2}$ |
| 1 | 0.5 | 4 | $-9.3 \times 10^{-2}$ | $2.8 \times 10^{-1}$ | $-1.6 \times 10^{-1}$ | $5.1 \times 10^{-1}$ |
| " | " | 5 | $-1.5 \times 10^{-2}$ | $1.9 \times 10^{-1}$ | $-6.6 \times 10^{-2}$ | $1.5 \times 10^{-1}$ |
| " | " | 6 | $-2.5 \times 10^{-2}$ | $1.4 \times 10^{-1}$ | $-1.5 \times 10^{-1}$ | $1.5 \times 10^{-1}$ |
| " | " | 7 | $-3.2 \times 10^{-2}$ | $3.9 \times 10^{-1}$ | **$3.1 \times 10^{1}$** | **$1.5 \times 10^{1}$** |
| " | 1 | 4 | $-7.5 \times 10^{-2}$ | $2.9 \times 10^{-1}$ | $-1.2 \times 10^{-1}$ | $5.0 \times 10^{-1}$ |
| " | " | 5 | $1.8 \times 10^{-2}$ | $7.4 \times 10^{-1}$ | $-8.9 \times 10^{-2}$ | $1.6 \times 10^{-1}$ |
| " | " | 6 | $-4.1 \times 10^{-2}$ | $2.9 \times 10^{-1}$ | $2.0 \times 10^{-1}$ | $3.1 \times 10^{-1}$ |
| " | " | 7 | $-2.4 \times 10^{-1}$ | $1.1 \times 10^{0}$ | **$7.6 \times 10^{1}$** | **$1.1 \times 10^{1}$** |
| 2 | 0.5 | 4 | $-2.4 \times 10^{-2}$ | $2.8 \times 10^{-1}$ | $-1.9 \times 10^{-1}$ | $5.4 \times 10^{-1}$ |
| " | " | 5 | $1.2 \times 10^{-2}$ | $9.5 \times 10^{-2}$ | $-1.0 \times 10^{-1}$ | $1.8 \times 10^{-1}$ |
| " | " | 6 | $-9.2 \times 10^{-3}$ | $6.6 \times 10^{-2}$ | $2.5 \times 10^{-1}$ | $1.5 \times 10^{0}$ |
| " | " | 7 | $-5.8 \times 10^{-3}$ | $3.2 \times 10^{-1}$ | **$2.5 \times 10^{1}$** | **$6.5 \times 10^{0}$** |
| " | 1 | 4 | $-7.5 \times 10^{-2}$ | $3.0 \times 10^{-1}$ | $-1.7 \times 10^{-1}$ | $4.7 \times 10^{-1}$ |
| " | " | 5 | $2.5 \times 10^{-2}$ | $7.7 \times 10^{-2}$ | $-6.8 \times 10^{-2}$ | $1.7 \times 10^{-1}$ |
| " | " | 6 | $1.2 \times 10^{-2}$ | $9.9 \times 10^{-2}$ | $2.3 \times 10^{-1}$ | $3.1 \times 10^{-1}$ |
| " | " | 7 | $-6.7 \times 10^{-2}$ | $5.0 \times 10^{-1}$ | **$4.5 \times 10^{1}$** | **$8.2 \times 10^{0}$** |
| 20 | 0.5 | 4 | $-3.7 \times 10^{-2}$ | $2.8 \times 10^{-1}$ | $-2.2 \times 10^{-1}$ | $3.7 \times 10^{-1}$ |
| " | " | 5 | $1.5 \times 10^{-2}$ | $8.8 \times 10^{-2}$ | $-5.0 \times 10^{-2}$ | $1.9 \times 10^{-1}$ |
| " | " | 6 | $-1.9 \times 10^{-2}$ | $1.4 \times 10^{-1}$ | $1.2 \times 10^{-1}$ | $2.0 \times 10^{-1}$ |
| " | " | 7 | $-2.3 \times 10^{-2}$ | $1.1 \times 10^{-1}$ | **$4.9 \times 10^{0}$** | **$2.2 \times 10^{-1}$** |
| " | 1 | 4 | $-6.9 \times 10^{-2}$ | $3.2 \times 10^{-1}$ | $-2.2 \times 10^{-1}$ | $3.7 \times 10^{-1}$ |
| " | " | 5 | $7.8 \times 10^{-3}$ | $1.5 \times 10^{-1}$ | $-7.3 \times 10^{-2}$ | $1.9 \times 10^{-1}$ |
| " | " | 6 | $-2.6 \times 10^{-2}$ | $2.2 \times 10^{-1}$ | $3.1 \times 10^{-1}$ | $3.8 \times 10^{-1}$ |
| " | " | 7 | $-2.8 \times 10^{-2}$ | $1.3 \times 10^{-1}$ | **$5.3 \times 10^{0}$** | **$7.9 \times 10^{-1}$** |
| 40 | 0.5 | 4 | $-5.3 \times 10^{-2}$ | $3.0 \times 10^{-1}$ | $-2.1 \times 10^{-1}$ | $3.7 \times 10^{-1}$ |
| " | " | 5 | $1.7 \times 10^{-2}$ | $8.4 \times 10^{-2}$ | $-7.6 \times 10^{-2}$ | $1.8 \times 10^{-1}$ |
| " | " | 6 | $1.9 \times 10^{-2}$ | $1.0 \times 10^{-1}$ | $9.9 \times 10^{-2}$ | $1.6 \times 10^{-1}$ |
| " | " | 7 | $1.5 \times 10^{-3}$ | $3.1 \times 10^{-3}$ | **$2.0 \times 10^{0}$** | **$2.9 \times 10^{-1}$** |
| " | 1 | 4 | $-5.6 \times 10^{-2}$ | $2.9 \times 10^{-1}$ | $-2.1 \times 10^{-1}$ | $3.9 \times 10^{-1}$ |
| " | " | 5 | $1.9 \times 10^{-2}$ | $8.9 \times 10^{-2}$ | $-6.1 \times 10^{-2}$ | $2.1 \times 10^{-1}$ |
| " | " | 6 | $-2.4 \times 10^{-2}$ | $7.6 \times 10^{-2}$ | $1.4 \times 10^{-1}$ | $2.1 \times 10^{-1}$ |
| " | " | 7 | $-3.4 \times 10^{-4}$ | $3.0 \times 10^{-2}$ | **$2.1 \times 10^{0}$** | **$7.6 \times 10^{-1}$** |

**Table 1**
Results for wavelet multiresolution analysis of CMB mapa in the ideal case of no instrumental noise. $F$, $\delta$ are, respectively, the void filling fraction and underdensity while $j$ is the analysis scale. $\sigma_S$ and $\sigma_K$ are the standard deviations of the skewness and kurtosis. Positive detections are highlighted with bold characters. The data for $F = \delta = 0\%$ are associated with Gaussian realisations.

The main observation here is that the kurtosis of the Haar wavelet expansion coefficients is capable of detecting the presence of voids on our CMB maps at the pixel scale with a margin of at least two standard deviations (depending on the specific values of the void parameters). On the other hand, the skewness would appear consistent with a Gaussian realisation for all maps analysed. This can be understood by taking into account the shape of the Haar wavelet (Figure 1). From symmetry considerations one can expect that there will be two subsets of expansion coefficients, *relative to the voids*, with similar absolute values, but opposite signs. Going back to Equation 18, we remark that the terms being summed are *odd* powers of the difference within brackets. Since there is roughly the same number of positive and negative coefficients, the terms within brackets will follow the same trend and the overall sum will be close to zero (mimicking a Gaussian distribution). The kurtosis, instead, can amplify the void signal since it is an *even* power and all terms in the sum will be positive.

Another relevant observation is that, as a general trend, the absolute value of the kurtosis (at $j = 7$) decreases as $F$ increases. This effect can be accounted for by considering that a smaller number of prominent features (such as voids) will give rise to a higher peakedness of the wavelet coefficient distribution (smaller number of coefficients describing the void signal) and, hence, to a higher value of the kurtosis. Also, higher values of $\delta$ are correlated with a larger kurtosis, as can be expected for stronger signals.

### 5.3. Results for maps with Planck-like instrumental noise levels.

We also studied the effect of simulated "Planck-like" instrumental noise on our results, such as whether instrumental effect could mask our previous findings. The predicted noise performance of the Planck-HFI instrument's 143 GHz channels was suited for our purpose. More specifically, we chose this frequency band since it is most sensitive one of HFI. In this case, the predicted value of the standard deviation for the estimated (zero–mean) Gaussian distributed noise is 2.2 μK/K, per pixel after 14 months of integration (ESA Technical Document, 2005). Our aim was not to simulate exactly the predicted Planck output maps but, rather, proof-of-concept "Planck-like" maps. The relevant noise was added to all maps examined in Section 5.2, without giving rise to clear visual differences with the original map set. The analysis was restricted to the kurtosis since, as we have seen, the skewness had proven ineffective. Our results are presented in Table 2.

| $F$ (%) | $\delta$ (%) | Scale $j$ | Kurtosis | $\sigma_K$ |
|---|---|---|---|---|
| 0 | 0 | 4 | $-1.5 \times 10^{-1}$ | $4.9 \times 10^{-1}$ |
| " | " | 5 | $-7.3 \times 10^{-3}$ | $1.8 \times 10^{-1}$ |
| " | " | 6 | $1.5 \times 10^{-2}$ | $1.5 \times 10^{-1}$ |
| " | " | 7 | $8.6 \times 10^{-3}$ | $8.1 \times 10^{-2}$ |
| 1 | 0.5 | 4 | $-1.7 \times 10^{-1}$ | $5.0 \times 10^{-1}$ |
| " | " | 5 | $-6.2 \times 10^{-2}$ | $1.8 \times 10^{-1}$ |
| " | " | 6 | $3.4 \times 10^{-2}$ | $1.8 \times 10^{-1}$ |
| " | " | 7 | **$4.4 \times 10^{0}$** | **$6.7 \times 10^{-1}$** |
| " | 1 | 4 | $-1.6 \times 10^{-1}$ | $5.6 \times 10^{-1}$ |
| " | " | 5 | $-6.7 \times 10^{-2}$ | $1.8 \times 10^{-1}$ |
| " | " | 6 | $1.3 \times 10^{-1}$ | $1.4 \times 10^{-1}$ |
| " | " | 7 | **$3.0 \times 10^{1}$** | **$7.7 \times 10^{0}$** |
| 2 | 0.5 | 4 | $-1.6 \times 10^{-1}$ | $5.1 \times 10^{-1}$ |
| " | " | 5 | $-7.1 \times 10^{-2}$ | $1.7 \times 10^{-1}$ |
| " | " | 6 | $5.9 \times 10^{-2}$ | $1.7 \times 10^{-1}$ |
| " | " | 7 | **$6.1 \times 10^{0}$** | **$8.0 \times 10^{-1}$** |
| " | 1 | 4 | $-1.7 \times 10^{-1}$ | $5.0 \times 10^{-1}$ |
| " | " | 5 | $-7.1 \times 10^{-1}$ | $1.7 \times 10^{-1}$ |
| " | " | 6 | $2.2 \times 10^{-1}$ | $4.9 \times 10^{-1}$ |
| " | " | 7 | **$2.6 \times 10^{1}$** | **$4.9 \times 10^{0}$** |
| 20 | 0.5 | 4 | $-2.0 \times 10^{-1}$ | $3.8 \times 10^{-1}$ |
| " | " | 5 | $-6.2 \times 10^{-2}$ | $1.8 \times 10^{-1}$ |
| " | " | 6 | $7.7 \times 10^{-2}$ | $1.0 \times 10^{-1}$ |
| " | " | 7 | **$1.7 \times 10^{0}$** | **$8.9 \times 10^{-2}$** |
| " | 1 | 4 | $-2.3 \times 10^{-1}$ | $4.3 \times 10^{-1}$ |
| " | " | 5 | $-4.9 \times 10^{-2}$ | $1.9 \times 10^{-1}$ |
| " | " | 6 | $2.7 \times 10^{-1}$ | $4.2 \times 10^{-1}$ |
| " | " | 7 | **$5.0 \times 10^{0}$** | **$1.8 \times 10^{-1}$** |
| 40 | 0.5 | 4 | $-2.0 \times 10^{-1}$ | $3.8 \times 10^{-1}$ |
| " | " | 5 | $-6.2 \times 10^{-2}$ | $1.8 \times 10^{-1}$ |
| " | " | 6 | $7.7 \times 10^{-2}$ | $1.0 \times 10^{-1}$ |
| " | " | 7 | **$1.7 \times 10^{0}$** | **$8.9 \times 10^{-2}$** |
| " | 1 | 4 | $-2.0 \times 10^{-1}$ | $3.8 \times 10^{-1}$ |
| " | " | 5 | $-5.0 \times 10^{-2}$ | $2.0 \times 10^{-1}$ |
| " | " | 6 | $1.1 \times 10^{-1}$ | $2.5 \times 10^{-1}$ |
| " | " | 7 | **$2.0 \times 10^{0}$** | **$2.9 \times 10^{-1}$** |

**Table 2**
Results for wavelet multiresolution analysis of CMB maps with predicted Planck-HFI 143 GHz channels' instrumental noise. $F$, $\delta$ are, respectively, the void filling fraction and underdensity while $j$ is the analysis scale. $\sigma_K$ is the standard deviation of the kurtosis. Positive detections are highlighted with bold characters. The data for $F = \delta = 0\%$ are associated with Gaussian realisations.

The evident result of adding noise on the maps is the marked decrease in the overall values of the kurtosis for the ΛCDM plus voids maps. Nevertheless, the void signatures are still clearly detectable through a statistical analysis. Since the Planck HFI 100 GHz channels have noise levels comparable to the 143 GHz ones, one could expect similar results also for this frequency band.

As an additional test we examined the situation with the estimated instrumental noise for the Planck HFI 353 GHz channel ($\sigma_{noise}$ = 14.7 μK/K after 14 months of integration). Strictly speaking, this is not a Planck CMB channel, but we were interested in investigating noise levels that were, approximately, an order of magnitude higher than those previously used. This kind of result could be useful for experiments other than Planck. We only considered the ΛCDM and $F$ = 1%, $\delta$ =1% cases. This combination of void parameters was chosen for a first test since, basing ourselves on the previous results, we expected it to give the highest value of the kurtosis in case of detection. Nevertheless, we found that at these levels, the instrumental noise completely masks the void signal and detection is not achievable.

## 6. Conclusions

In this paper we discuss the potential detection of traces that voids from first-order phase transitions might imprint on the Cosmic Microwave Background. These voids have a twofold importance, since their detection could provide a direct verification of inflationary theory (this being the main scientific motivation for their detection). They have also been proposed as alternative (or complementary) seeds for large-scale structure formation. We apply the technique of wavelet analysis to a set of simulated CMB maps with a simplified void model and varying void parameter values. These were chosen so that the voids would not be detectable from a power spectrum analysis alone. We find that the kurtosis of the Haar wavelet expansion coefficients is able to detect these void pixels in maps without instrumental noise. Moreover, detection is still possible when Planck-like (HFI 143 GHz channels) simulated Gaussian noise is added to the maps. This was no longer true when the noise increased by about an order of magnitude, such as in the Planck HFI 353 GHz channels.

We envisage that this kind of work could be furthered, for instance by considering CMB maps exhibiting void signatures with different sizes. This would probably require the use of different wavelet bases for the detection of pixels-sized voids as opposed to larger ones, due to their different shapes on the CMB map. Although we show that the wavelet technique gave promising results, it is not advisable to rely on a single method to detect the presence of such weak signals and further work to make available a combination of methods would be advisable. This also in view of discriminating between different early universe signatures and astrophysical foregrounds.